\documentclass[11pt,oneside]{article}

\RequirePackage{fix-cm}

\usepackage[operators,sets]{cryptocode}

\newcommand{\C}{\mathbb{C}}
\usepackage{authblk}

\begin{document}

\title{Concerning Quantum Identity Authentication Without Entanglement}
\author[1]{Carlos E. Gonz\'alez-Guill\'en\thanks{carlos.gguillen@upm.es}}
\affil[1]{Departamento de Matem\'atica Aplicada a la Ingenier\'ia Industrial, Universidad Polit\'ecnica de Madrid, Spain}
\author[2]{Mar\'{\i}a Isabel Gonz\'alez Vasco\thanks{mariaisabel.vasco@urjc.es}}
\affil[2]{MACIMTE, Universidad Rey Juan Carlos, Madrid, Spain.}
\author[3]{Floyd Johnson\thanks{johnsonf2017@fau.edu}}
\affil[3]{Florida Atlantic University, Boca Raton, Florida, United States of America}
\author[2]{\'Angel L. P\'erez del Pozo\thanks{angel.perez@urjc.es}}

\date{}

\maketitle

\begin{abstract}
Identification schemes are interactive protocols typically involving two parties, a \emph{prover}, who wants to provide evidence of his or her identity and a  \emph{verifier}, who checks the provided evidence and decide whether it comes or not from the intended prover. 

In this paper, we comment on a recent proposal for quantum identity authentication from Zawadzki~\cite{Zawadzki19},  and give a concrete attack upholding theoretical impossibility results from Lo~\cite{Lo97} and Buhrman et al.~\cite{Buhrman12}. More precisely, we show that using a simple strategy an adversary may indeed  obtain non-negligible information on the  shared identification secret. While the security of a quantum identity authentication scheme is not formally defined in \cite{Zawadzki19}, it is clear that such a definition should somehow imply that an external entity may gain no information on the shared identification scheme (even if he actively participates injecting messages in  a protocol execution, which is not assumed in our attack strategy).

\end{abstract}

\section{Introduction}
One of the major goals of cryptography is authentication in different flavours, namely, providing guarantees that certain interaction is actually involving certain parties from a designated presumed set of users.  In the two party scenario, cryptographic constructions towards this goal are called \emph{identity authentication schemes}, and have been extensively studied in classical cryptography.  The advent of quantum computers spells the possible end for many of these protocols however.  

Since Wiesner proposed using quantum mechanics in cryptography in the 1970's multiple directions using this concept have undergone serious research.  One major role quantum mechanics has played in cryptography is the development of quantum key distribution (QKD) where two parties can securely share a one time pad using quantum mechanics, for example the seminal protocol BB84 \cite{BB84}.  One drawback most of these protocols share is the need for authentication, which is traditionally done over an authenticated classical channel.  

Classically, there are different ways of defining so-called \emph{identification schemes}, for mutual authentication of peers, mainly depending on whether the involved parties share some secret information (such as a password) or should rely on different (often certified) keys provided by a trusted third party.  In the quantum scenario, different identification protocols have been introduced following the first approach, e.g., assuming that two parties may obtain authentication evidence from the common knowledge of a shared secret. These kind of constructions, often called  \emph{quantum identity authentication schemes} (or just \emph{quantum identification schemes}), are thus closely related to protocols for \emph{quantum equality tests} and \emph{quantum private comparison}.  All these constructions are concrete examples of two-party computations with asymmetric output, i.e. allowing only one of the two parties involved to learn the result of a computation on two inputs. Without posing restrictions on an adversary it was shown by Lo in \cite{Lo97} and and Buhrman et al. in~\cite{Buhrman12} that these constructions are impossible, even in a quantum setting. As a consequence, constructions for generic unrestricted adversaries in the quantum setting are doomed to failure.

All in all,  the necessity for authentication in QKD has led to many authors considering approaches  which are strictly quantum in nature, such as those in \cite{Penghao16,Zeng00,Huang11} which are based off entanglement or more recently \cite{Zawadzki19,Hong17} which do not rely on entanglement.  These are known as \emph{quantum identity authentication} (QIA) protocols.  For protocols such as BB84 that do not rely on entanglement it would be more appealing to not rely on entanglement for entity authentication purposes. 

\medskip

\noindent{\it Our Contribution.} Recently, an original work about authentication without entanglement by Hong et. al. in \cite{Hong17} was improved by Zawadzki using tools from classical cryptography in \cite{Zawadzki19}.  We start this contribution by summarizing in section~\ref{sec:impossibility} the  impossibility results from  Lo~\cite{Lo97} and Buhrman et al.~\cite{Buhrman12}, concerning generic quantum two party protocols. Further, we present and discuss the Zawadzki protocol in section \ref{sec:zawadzki_protocol} and show how it succumbs  under a simple attack, which we outline in section \ref{sec:attack}.  Our attack evidences the practical implications of the proven impossibility of identification schemes  as conceived in Zawadki's design,  and thus we stress that fundamental changes in the original proposal, beyond preventing our attack, would be needed in order to derive a secure identification scheme.

\section{Quantum  Equality Tests are Impossible}\label{sec:impossibility}

A \emph{one sided equality test} is a cryptographic protocol in which one party, Alice, convinces another, Bob, that they share a common key by revealing nothing to either party except equality (or inequality) to Bob.  Formally we define a key space $K$ and a function $F:K^2\to \{0,1\}$ which checks for equality.  Let $i\in K$ be Alice's key and $j\in K$ be Bob's key.  The goals of a one sided equality test are as follows: 

1) $F(i,j)=1$ if and only if $i=j$.

2) Alice learns nothing about $j$ nor about $F(i,j)$.

3) Bob learns $F(i,j)$ with certainty.  If $F(i,j)=0$ then Bob learns nothing about $i$ except $i\neq j$.
\newline
 The above is a specific case of a one-sided two-party secure computation protocol as described in \cite{Lo97}. In this work, a very general result is proven indicating that any protocol realising a one-sided two party secure computation task is impossible, even in a quantum setting. In particular, Lo shows in~\cite{Lo97} that if a protocol satisfies 1) and 2) then Bob can know the output of $F(i,j)$ for any $j$. Furthermore, the one sided equality test with some small relaxations on points 1) and 3) is also proven impossible. Hence, any one-sided QIA protocol which validates identities using equality tests by use of quantum mechanics is impossible without imposing restrictions on an adversary.

Note that the above argument  says nothing about protocols with built in adversarial assumptions such as those presented in \cite{Damgard14,Bouman13}. Further, note that many of the QIA schemes end up with a round where Bob accepts or rejects, which makes Alice aware of the success or failure of the protocol. Indeed,  those schemes can be straightforwardly turned into one-sided equality tests by suppressing Bob's final message announcing the result.  Hence, they are clearly insecure against a dishonest Bob. However, note that if any such protocol can be modified so that  Alice may obtain information on the identification output at some point before the last protocol round, it is unclear how Lo's impossibility result would apply. However, if they are built upon equality tests we can get impossibility from another well know result by Buhrman el al.\cite{Buhrman12}.  Certainly,  two-sided QIA schemes, in which both Alice and Bob learn the result of the protocol, are a particular case of two-sided two-party computations. It is shown in~\cite{Buhrman12} that a correct quantum protocol for a classical two-sided two-party computation that is secure against one of the parties is completely insecure against the other. For equality tests, if one of the parties, say Alice, learns nothing else than $F(i,j)$, the other party, Bob, will indeed be able to compute  $F(i,j)$ for all possible inputs $j$. Thus, any two-sided QIA protocol which validates identities using equality tests is also impossible without imposing further restrictions on the adversary. 

\section{QIA without Entanglement}\label{sec:zawadzki_protocol}
Here we will outline the protocol proposed in \cite{Zawadzki19} with some minor modifications, discussed afterword.  Suppose Alice and Bob have keys $k_a$ and $k_b$ respectively.  Bob wishes to verify that $k_b=k_a$ without leaking any information about $k_b$ or $k_a$.  Bob randomly generates a nonce $r$ from a designated domain and generates a universal hash function $H:\{0,1\}^N\to \{0,1\}^{2d}$.  This hash function may be chosen by Bob or sampled at random, in the below description we sample from a space of universal hash functions with image $\{0,1\}^{2d}$ called $\mathbb{H}$.  Bob sends Alice $r$ and $H$.  Alice then calculates the value $h_a=H(r||k_a)$.  Alice then acts on pairs in $h_a$ with an embedding function $Q:\{0,1\}^2\to \CC ^2$.  This function $Q$ uses the first of the two binary values to determine the measurement basis (horizontal/vertical or diagonal/antidiagonal) and the second to determine the specific qubit in $\{|0\rangle, |1\rangle , |+\rangle, |-\rangle\}$.  For example, $Q(0,0)=|0\rangle$ and $Q(1,0)=|+\rangle$.  Applying $Q$ to the pairs of bits in $h_a$ Alice prepares and sends $d$ qubits to Bob over the quantum channel.

Upon reception, Bob computes $h_b=H(r||k_b)$.  Note that if $k_a=k_b$ then $h_a=h_b$.  Using the first bit of each pair Bob measures the quantum states and insures he obtains the correct qubit according to the second bit of the pair.  If Bob measures something that disagrees with the even bits of $h_b$ then Bob rejects Alice's challenge.  If after measuring all qubits Bob has not yet rejected Alice's challenge then he accepts her challenge.

Changes made to the protocol are as follows: (1) Bob generates $r$ and $H$, this is done to thwart a simple attack discussed later; (2) the hash function changes between trials, this has no impact on the security of the protocol due to the public nature of the hash in both instances; and finally (3) here we assume for simplicity that Alice and Bob obtain the same nonce $r$ with certainty, using classical error correction techniques one can be relatively certain both parties obtain the same nonce.  See below for a schematic overview of the protocol.

 \begin{center}
\pseudocode{%
\textbf{Alice}\< \< \textbf{Bob} \\ [0.1\baselineskip][\hline]
\<\< \\
k_a \< \< k_b \\
\<\<r \sample \{0,1\}^* \\
 \<\< H \sample \mathbb{H} \\
\< \sendmessageleft{top=$r$ $H$, bottom=Classical} \< \\
   h_a \gets H(r||k_a) \<\< \\
  |\varphi _i \rangle \gets Q(h_{a_{2i-1}},h_{a_{2i}})\<\< \\
  \text{repeat for all i=1,2,...,d} \<\< \\
   \<\< h_b\gets H(r||k_b) \\
  \< \sendmessageright{top=$|\varphi _i\rangle$  $\forall i\leq d$, bottom=Quantum} \< \\
 \<\< s_i\gets M(|\varphi _i \rangle, h_{b_{2i-1}}) \\
 \<\<\text{repeat for all i=1,2,...,d} \\
 \<\< \text{if $s_i = h_{b_{2i}} \forall i\leq d$, accept}\\
 \<\< \text{otherwise, reject}\\
 \< \sendmessageleft{top=Accept/Reject, bottom=Classical} \< \\
 \\ [0.1\baselineskip][\hline]}
%\hline 
 $\quad$\\
 \centerline{Figure 1. The protocol presented in~\cite{Zawadzki19}}
\end{center}

 The reason we force Bob to generate the randomness instead of Alice is that an adversary with unbound quantum memory may impersonate Bob but not make a measurement.  Suppose an adversary does not know the key but requests Alice to identify herself.  If Alice generates and sends $r,H$ with the string of states $|\varphi _i \rangle$ then the adversary may record $r,H$ and hold in memory, but not measure, the qubits.  At a later time an honest participant may ask the adversary to identify themselves, in this case the adversary may send $r,H$ and the qubits in memory.  Thus, the adversary correctly forges an authentication.  Note that as we have presented the algorithm an adversary may still make this impersonation by waiting between Alice and Bob then passing the information between the two.  The difference is as long as Bob generates the nonce then this attack must only be done while Alice and Bob are both online, whereas if Alice generates and sends the nonce then an adversary may hold the states for as long as is technologically feasible.

 The proposed protocol is claimed to be leakage resistant when considering an adversary measuring in a random basis.  The reasoning behind this is that unless an adversary, Eve, correctly guesses the correct basis for each round, she will obtain different values for at least one of the bits of the hash.  Now suppose an adversary is capable of computing preimages of hash functions through brute force with unbounded classical computational power or through dictionary attacks with unbounded classical memory.  In this case it is unlikely that there will exist a $k_e\in K$ such that $H(r||k_e)$ matches what Eve measured.  In the event there does exist such a $k_e$ then with overwhelming probability $k_e\neq k_a=k_b$ and Eve will not be able to falsify authentication of Alice or Bob. 
 
Unfortunately, the proposed protocol is claimed to be exactly a two-sided equality test with possible, though unlikely, relaxation of $F(i,j)=1$ if and only if $i=j$ (in this case $i$ is $k_a$ and $j$ is $k_b$).  We know such a protocol has necessary leakage and due to the non-interactive nature of Bob we know that $k_b$ has no leakage, thus we know there must exist some leakage on $k_a$.  Although Eve may not be able to determine any exact bit of $k_a$ she may drastically reduce the number of possible options for $k_a$ and hence construct a proper subset of $K$ such that the true value for $k_a$ is contained in this subset.  An attack exemplifying this phenomenon is described in the next section.

 \section{A Key Space Reduction Attack on QIA without Entanglement (our contribution)}\label{sec:attack}
 Before discussing the specific attack, let $B$ be a set of orthogonal bases in $\C^2$ and consider the following fact.  If a quantum state is prepared in a basis $b\in B$ with value $v\in \{0,1\}$, then an adversary may always remove one possible combination of $b$ and $v$ with a single measurement. Upon measuring in basis $b'\in B$ an adversary obtains $v'\in \{0,1\}$.  The adversary is then certain the original pair $(b,v)$ was not $(b',1\bigoplus v')$, as when measured in the basis $b$ the qubit prepared by $b$ and $v$ will yield $v$ with certainty. Note that the adversary cannot say with certainty how the qubit was prepared, but they can always remove one possible option.

  Suppose now that instead of sampling at random for $b$ and $v$, the qubit is prepared using a private key $k\in K$ and a set of public parameters $p$, namely $b=b(k,p)$ and $v=v(k,p)$.  An adversary once again measures in basis $b'\in B$ (chosen or taken at random) to obtain $v'\in \{0,1\}$, they may then determine a basis/value pair in which the qubit was not prepared.  Because the adversary is assumed to be computationally unbounded they may then compute $b(\hat{k},p)$ and $v(\hat{k},p)$ for all $\hat{k}\in K$.  Whenever these computations output the impossible pair $k',v'$ the adversary becomes aware that $\hat{k}\neq k$, hence reducing the key space.  The extent to which the key space is reduced depends on the number of basis in $B$.   If the distribution of basis choices in $B$ is low entropy the attack may be accomplished as described while if $B$ is high entropy then a probabilistic version decreases the space of likely keys.  The assumption that the adversary is computationally unbounded may be lifted if $k$ is low entropy (for he can then indeed test all possible values for $k$ --- given there are only a polynomial set of candidates), however assuming a computationally bounded adversary immediately removes unconditional security as an end goal.

 Let us now apply this key space reduction to the QIA protocol proposed in \cite{Zawadzki19}, in this case the private key is $k$ and the public parameters are $r$ and $H$.  
 Suppose an Eve has no a priori knowledge of the key except its existence in $K$.  After receiving $r$ and $H$ over the classical channel she measures all qubits $|\varphi _i \rangle$ received from Alice in the horizontal/vertical basis and records the outputs as $M$.  In the case where Eve is utilizing man-in-the-middle she is done, if she is impersonating Bob she accepts or rejects the protocol.

 After the protocol finishes the adversary may then compute $h_{\hat{k}}=H(r||\hat{k})$ for all $\hat{k}\in K$. Suppose the first qubit Eve measured in $M$ was $|0\rangle$.  She now examines the first two bits of each $h_{\hat{k}}$, those that begin 00, 10, or 11 are all possible of obtaining the qubit $|0\rangle$ after measurement.  The first of these three tuples will yield $|0\rangle$ with certainty and the later two with a probability of 0.5.  The final tuple 01 however is not possible as that would imply that the qubit started in the state $|1\rangle$ and measured in $|0\rangle$.  Thus, Eve knows that any $\hat{k}$ such that $h_{\hat{k}}$ begins 01 is not the key.  The hash function is assumed to be independent and identically distributed so this removes approximately $\frac{1}{4}$ of all possible keys.  Repeat this process for all qubits.  After completion of all hash and check operations the adversary has obtained a subset of the key space which contains the key, hence causing information leakage.  Specifically, the adversary knows the key is in subset $S$ defined by $$S=\{s\in K: h_{s_{2i}}=M_{i} \text{ and } h_{s_{2i-1}}=0\ \forall i\leq d\}.$$  Note that the true key $k\in S$ and $|S|\approx (\frac{3}{4})^d |K|$.  This attack may be repeated using other choice of bases (i.e.- not always selecting the horizontal/vertical basis) and utilizing the same approach these different bases will likely yield different subsets of $K$.  The key is in the intersection of all these subsets, decreasing the possible key space further.  Note that after applying this attack the advantage of adversary may be negligible yet if $(\frac{3}{4})^d |K|$ is still not negligible in the security parameter.  Parameters are not listed in \cite{Zawadzki19} however it does not seem unreasonable that $d$ is sufficiently large compared to $|K|$, otherwise a false positive for authentication is more likely.
 
\section{Other QIA protocols} 

It is worth pointing out that the attack described in section \ref{sec:attack} also applies to the protocol by Hong et al. \cite{Hong17}, which Zawadzki \cite{Zawadzki19} modifies. In more detail, the protocol in \cite{Hong17} is similar to Zawadzki's, but do not use a hash function. Instead, whenever Alice transmits the qubits sequentially and, before sending each qubit, she randomly decides if she is going to use \emph{security mode} or \emph{authentication mode}. In the first case, she sends a decoy state while in the second one, a qubit encoding two bits of the authentication string is sent, similarly to \cite{Zawadzki19}. After Bob's reception, Alice announces which mode she just has used. Therefore an adversary using the same strategy described in our attack in section \ref{sec:attack} and collecting the information obtained whenever Alice announces authentication mode, will be able to shrink the size of the key space in the same way we have previously stated.

On the other hand, other quantum identification protocols proposed in the literature are not vulnerable to our attack neither contradict the impossibility result mentioned in section \ref{sec:impossibility}. For instance, some of them \cite{Penghao16,Zeng00,Yang13} are aided by the presence of a trusted third party, therefore not being real two-party protocols. Another type of protocols, such as \cite{Mihara02,Shi01,Zhang00}, make use of an entangled quantum state shared between both parties. In \cite{Mihara02} the users, in addition, share a bitstring used as a password; both parties measures their part of the entangled state to produce a one time key that one of the users XORs with the password and sends the result to the other who checks for consistency. The downside of this approach is that, to repeat the identification process, the parties need to be provided again with new entangled states. In \cite{Shi01,Zhang00} the users do not share any classical secret, they just use the entangled state to identify themselves.

 \section{Conclusion}
The protocol given by Zawadzki in \cite{Zawadzki19} may be secure against hash preimage attacks when attempting to find an exact match, however when considering impossible results from quantum measurements we see some hashed key values are not possible.  Proverbially, the forest may be secure but each of the trees reveals enough information to reconstruct the possible forests.  By eliminating approximately one quarter of the key options from each qubit we see that by measuring all the individual qubits in a random basis does in fact reveal a great deal about the key.  This attack has no concern on quantum memory though relies heavily on classical computation power.  Hence, unlike \cite{Damgard14,Bouman13} where the authors consider a bounded quantum storage model, the only way to make this protocol secure without greatly changing its construction is to constrict an adversaries computational power.

The attack proposed here is general in the sense of QIA protocols in the prepare and measure setup, thus any future protocol of this type must consider possible key space reduction attacks.  Regardless of the method it is known that any identification protocol which poses no bounds on the adversary will inevitably fail due to results of Lo and Buhrman et al.  For this reason we advise that any future attempts at identification schemes consider, and clearly communicate, their assumptions and objectives.

\vspace{2em}\noindent{\bf {Acknowledgements:}} 
This research was sponsored in part by the NATO Science for Peace and Security Programme under grant G5448, in part by Spanish MINECO under grants MTM2016-77213-R and MTM2017-88385-P, and in part by Programa Propio de I+D+i of the Universidad Polit\'ecnica de Madrid.

\bibliographystyle{unsrt}
\bibliography{QIA.bib}

\begin{thebibliography}{10}

\bibitem{Zawadzki19}
Piotr Zawadzki.
\newblock Quantum identity authentication without entanglement.
\newblock {\em Quantum Information Processing}, 18(1):7, 2019.

\bibitem{Lo97}
Hoi-Kwong Lo.
\newblock Insecurity of quantum secure computations.
\newblock {\em Physical Review A}, 56(2):1154, 1997.

\bibitem{Buhrman12}
Harry Buhrman, Matthias Christandl, and Christian Schaffner.
\newblock Complete insecurity of quantum protocols for classical two-party
  computation.
\newblock {\em Phys. Rev. Lett.}, 109:160501, 2012.

\bibitem{BB84}
Charles~H Bennett and Gilles Brassard.
\newblock Quantum cryptography: public key distribution and coin tossing.
\newblock {\em Theor. Comput. Sci.}, 560(12):7--11, 2014.

\bibitem{Penghao16}
Niu Penghao, Chen Yuan, and Li~Chong.
\newblock Quantum authentication scheme based on entanglement swapping.
\newblock {\em International Journal of Theoretical Physics}, 55(1):302--312,
  2016.

\bibitem{Zeng00}
Guihua Zeng and Weiping Zhang.
\newblock Identity verification in quantum key distribution.
\newblock {\em Physical Review A}, 61(2):022303, 2000.

\bibitem{Huang11}
Peng Huang, JUN Zhu, Yuan Lu, and Gui-Hua Zeng.
\newblock Quantum identity authentication using gaussian-modulated squeezed
  states.
\newblock {\em International Journal of Quantum Information}, 9(02):701--721,
  2011.

\bibitem{Hong17}
Chang ho~Hong, Jino Heo, Jin~Gak Jang, and Daesung Kwon.
\newblock Quantum identity authentication with single photon.
\newblock {\em Quantum Information Processing}, 16(10):236, 2017.

\bibitem{Damgard14}
Ivan~B Damg{\aa}rd, Serge Fehr, Louis Salvail, and Christian Schaffner.
\newblock Secure identification and qkd in the bounded-quantum-storage model.
\newblock In {\em Annual International Cryptology Conference}, pages 342--359.
  Springer, 2007.

\bibitem{Bouman13}
Niek~J. Bouman, Serge Fehr, Carlos Gonz{\'a}lez-Guill{\'e}n, and Christian
  Schaffner.
\newblock An all-but-one entropic uncertainty relation, and application to
  password-based identification.
\newblock In {\em Theory of Quantum Computation, Communication, and
  Cryptography}, pages 29--44. Springer, 2013.

\bibitem{Yang13}
Yu-Guang Yang, Hong-Yang Wang, Xin Jia, and Hua Zhang.
\newblock A quantum protocol for (t, n)-threshold identity authentication based
  on greenberger-horne-zeilinger states.
\newblock {\em International Journal of Theoretical Physics}, 52(2):524--530,
  2013.

\bibitem{Mihara02}
Takashi Mihara.
\newblock Quantum identification schemes with entanglements.
\newblock {\em Physical review A}, 65(5):052326, 2002.

\bibitem{Shi01}
Bao-Sen Shi, Jian Li, Jin-Ming Liu, Xiao-Feng Fan, and Guang-Can Guo.
\newblock Quantum key distribution and quantum authentication based on
  entangled state.
\newblock {\em Physics letters A}, 281(2-3):83--87, 2001.

\bibitem{Zhang00}
Yong-Sheng Zhang, Chuan-Feng Li, and Guang-Can Guo.
\newblock Quantum authentication using entangled state.
\newblock {\em arXiv preprint quant-ph/0008044}, 2000.

\end{thebibliography}

\end{document}